\documentclass[twocolumn,showpacs,preprintnumbers,amsmath,amssymb]{revtex4}
\usepackage{tabularx,graphicx}

\usepackage{color}
\usepackage{hyperref}
\hypersetup{
    colorlinks=true,
    linkcolor=blue,
    filecolor=blue,      
    urlcolor=blue,
}

\usepackage{color}

\usepackage{ulem}   % to strike things out
\begin{document}
%\documentstyle[aps]{revtex}
%\documentstyle[preprint,aps]{revtex}
%\begin{document}

\newcommand{\beq}{\begin{equation}}
\newcommand{\eeq}{\end{equation}}
\newcommand{\beqn}{\begin{eqnarray}}
\newcommand{\eeqn}{\end{eqnarray}}
\newcommand{\bmath}{\begin{subequations}}
\newcommand{\emath}{\end{subequations}}
\newcommand{\bra}[1]{\langle #1|}
\newcommand{\ket}[1]{|#1\rangle}

%\draft
\title{Absence of magnetic evidence for superconductivity in hydrides under high pressure}

\author{J. E. Hirsch$^{a}$  and F. Marsiglio$^{b}$ }
\address{$^{a}$Department of Physics, University of California, San Diego,
La Jolla, CA 92093-0319\\
$^{b}$Department of Physics, University of Alberta, Edmonton,
Alberta, Canada T6G 2E1}

\begin{abstract} 
It is generally believed that magnetization measurements on sulfur hydride under high pressure
performed in 2015 \cite{sh3}  provided ``final convincing evidence of superconductivity'' \cite{review1} in that material,
in agreement with theoretical predictions \cite{th0,th1}. Supported by this precedent,  drops in resistance that were later observed
in several  other hydrides under high pressure \cite{review1,review2} have been generally accepted as evidence of superconductivity without corroborating
evidence from magnetic measurements. In this paper we challenge the original interpretation that 
the magnetic measurements on sulfur hydride performed in 2015 were evidence of superconductivity. 
We  point out that a large $paramagnetic$ contribution to the magnetic susceptibility was detected  below $T_c$
and argue that  its temperature dependence rules out the possibility that it would be a background signal; instead  the temperature
dependence
indicates  that  the paramagnetic behavior originated in the  sample.  We discuss possible explanations for this remarkable behavior and conclude that standard superconductors would not show such behavior. 
We also survey all the other published data from magnetic measurements on this class of materials and conclude that they do not provide
strong evidence for superconductivity. Consequently, we call into question the generally accepted view
that conventional superconductivity in hydrogen-rich materials at high temperature and pressure is a reality,
and discuss the implications if it is  not.   
\end{abstract}
\pacs{}
\maketitle 
\section{introduction}

High-temperature conventional superconductivity has reportedly been found   in several hydrogen-rich materials under high pressure
in recent years \cite{review1,review2}. This behavior is predicted by calculations based on the conventional
 theory of superconductivity \cite{review1,review2,theory1,theory2}. 
The first material where this phenomenon was reportedly observed was sulfur hydride under pressures above 100GPa \cite{sh3},
reaching a critical temperature of  203K for pressure 155GPa. In this paper we analyze the magnetic
evidence on which that claim was based. We argue  that it indicates that the material is either a
non-standard superconductor \cite{hm,hm2,hm3} or more likely is not a superconductor.

More generally, in recent work we \cite{hm,hm2} and others \cite{dc} have called into question the interpretation that resistance drops measured in several  high pressure hydrides are
evidence of superconductivity. In a carbonaceous sulfur hydride claimed to become superconducting at room
temperature \cite{roomt}, the resistance versus temperature data show extremely sharp drops, uncharacteristic of  high-temperature superconductivity \cite{hm,dc}.
Even more unusual  is the fact that under application of a magnetic field the  drops in resistance  remain equally sharp.
That is not the behaviour expected from standard superconductors, whether conventional or unconventional \cite{tinkham}. 
We furthermore pointed out that similar unphysical behaviour is observed in several other hydrides claimed to 
become superconducting under high pressure. We proposed  a new category of `nonstandard superconductors'
to describe superconducting materials with such unusual behavior,   and raised the possibility that the behavior observed  is not due to  superconductivity \cite{hm2}.

In this paper we focus on magnetic evidence of superconductivity in hydrides under high pressure. We start by analyzing the claim that the magnetization data for sulfur hydride
reported in 2015 \cite{sh3} demonstrated that sulfur hydride under high pressure is a high-temperature superconductor.

\section{magnetization of sulfur hydride}

Figure 1 shows the magnetization versus magnetic field data reported in Ref.~\cite{sh3} (see also \cite{drozdov}) from which it is
inferred that the material is a superconductor below $203$~K. It is argued in Refs.~\cite{sh3,drozdov} that the graphs show the diamagnetic response of
the superconductor superposed to a paramagnetic signal originating in the diamond anvil cell (DAC) where the sample resides.

However, from the data shown in Fig.~1 we can extract the paramagnetic contribution to the magnetization. We plot the
corresponding magnetic susceptibility $\chi=M/H$, with M the magnetization and H the applied field) versus temperature in
Fig.~2.

%figure 1
\begin{figure}[h]
 \resizebox{4.5cm}{!}{\includegraphics[width=6cm]{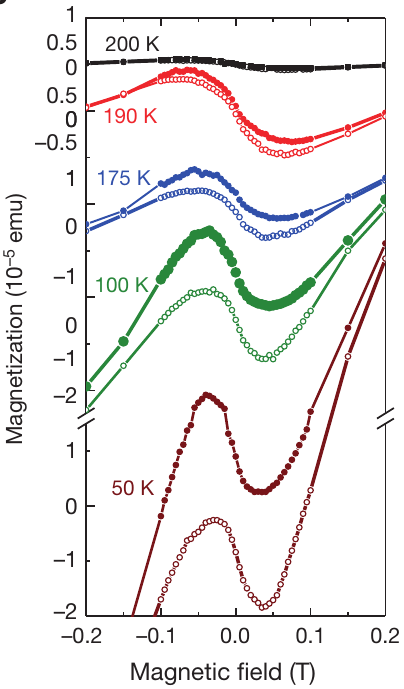}} 
\caption{Magnetization measurements for sulfur hydride at a pressure of 155GPa from Fig.~4c of ref. \cite{sh3}.
The open (closed) circles correspond to magnetic field increasing (decreasing).
}
\label{fig3}
\end{figure}

Fig.~2 shows that the paramagnetic contribution goes to zero, or very close to zero, at or around the inferred superconducting
critical temperature of the sample, $203$~K. The three data points shown in the figure at temperatures above $200$~K were obtained from 
Fig.~4d of Ref.~\cite{sh3}.
If the paramagnetic contribution resulted from a magnetic background signal in the DAC, we would expect it to follow a Curie-Weiss or Curie law.
In Fig.~2 we have fitted such laws to the measured values of the paramagnetic susceptibility at $50$~K and $200$~K. It can be seen
that neither of them fits the observed behavior. 

%figure 2
\begin{figure}[h]
 \resizebox{8.5cm}{!}{\includegraphics[width=6cm]{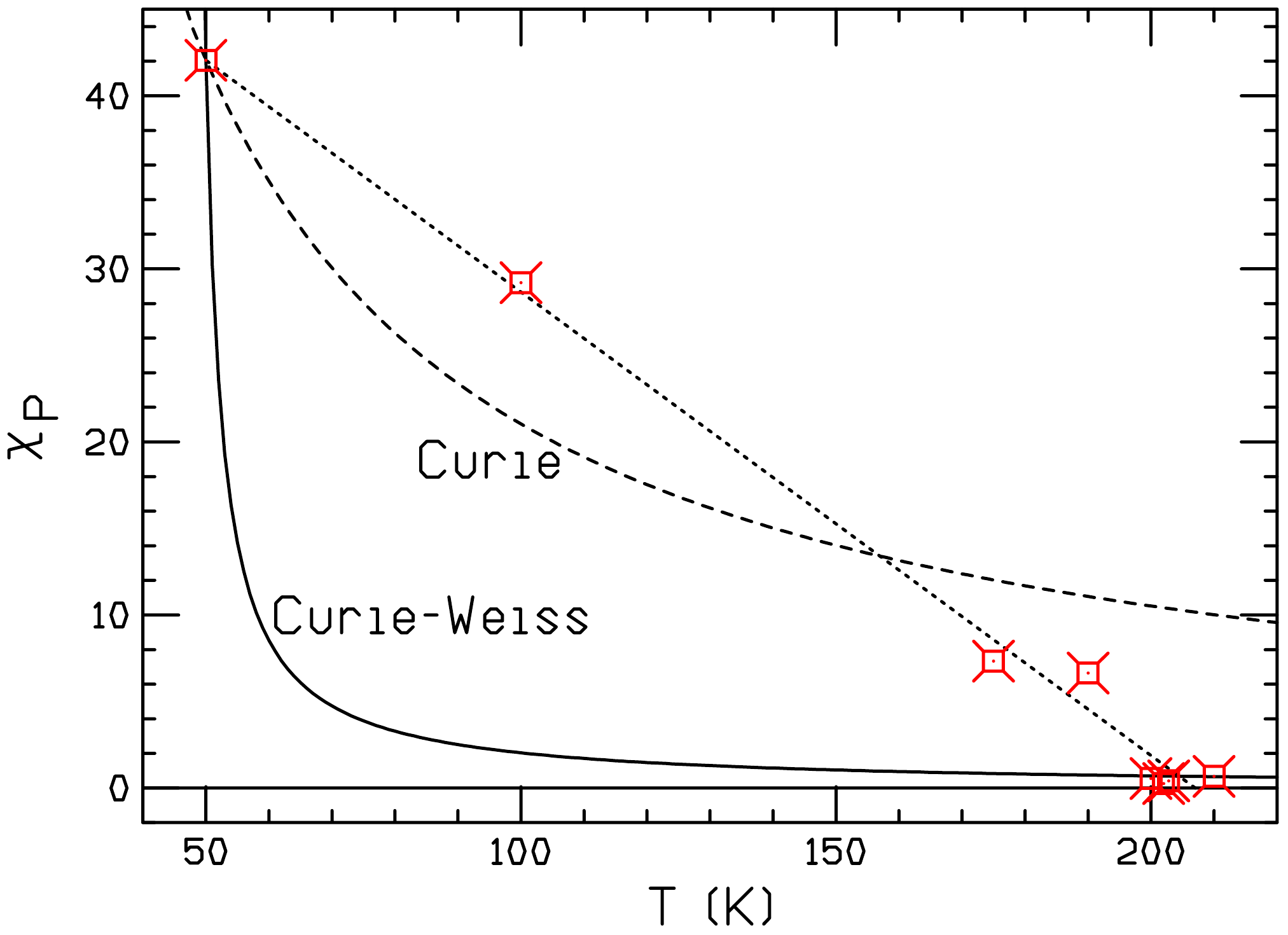}} 
\caption{Paramagnetic contribution to the susceptibility $\chi_P$ inferred from Fig. 1, in units $10^{-5}$~emu/T. 
The experimental data are the points, the dotted line is a guide to the eye.
}
\label{fig3}
\end{figure}

Let us consider for comparison similar data obtained for the superconductor $GdBa_2Cu_3O_7$ in Ref.~\cite{ref31}, shown in Fig.~3.
This paper was cited in Ref.~\cite{sh3} as giving analogous behavior to the one shown in Fig.~1 due to the presence of the
magnetic $Gd$ atoms.  

%figure 3
 \begin{figure}[h]
 \resizebox{5.9cm}{!}{\includegraphics[width=6cm]{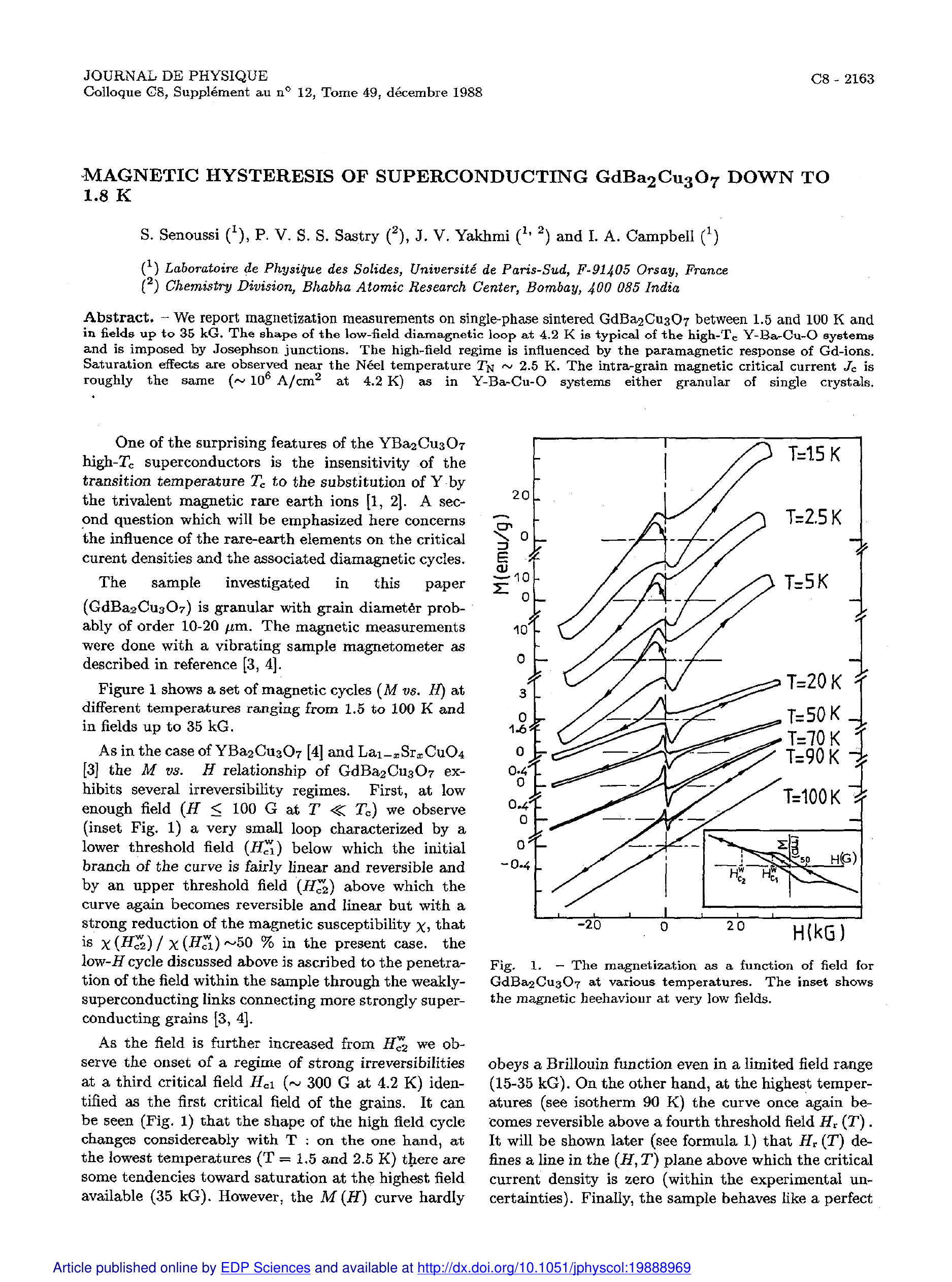}} 
\caption{Magnetization measurements for $GdBa_2Cu_3O_7$   from Fig. 1 of Ref.~\cite{ref31}.
}
\label{fig3}
\end{figure}

Indeed, we see a similarity in the magnetization data of Fig.~3 and Fig.~1: a diamagnetic signal for small fields is
superposed with a paramagnetic signal that persists to much larger fields. Following the same procedure as in obtaining
Fig.~2 from Fig.~1, we show in Fig.~4 the paramagnetic contribution to the susceptibility inferred from the data in Fig.~3.

%figure 4
\begin{figure}[h]
 \resizebox{8.5cm}{!}{\includegraphics[width=6cm]{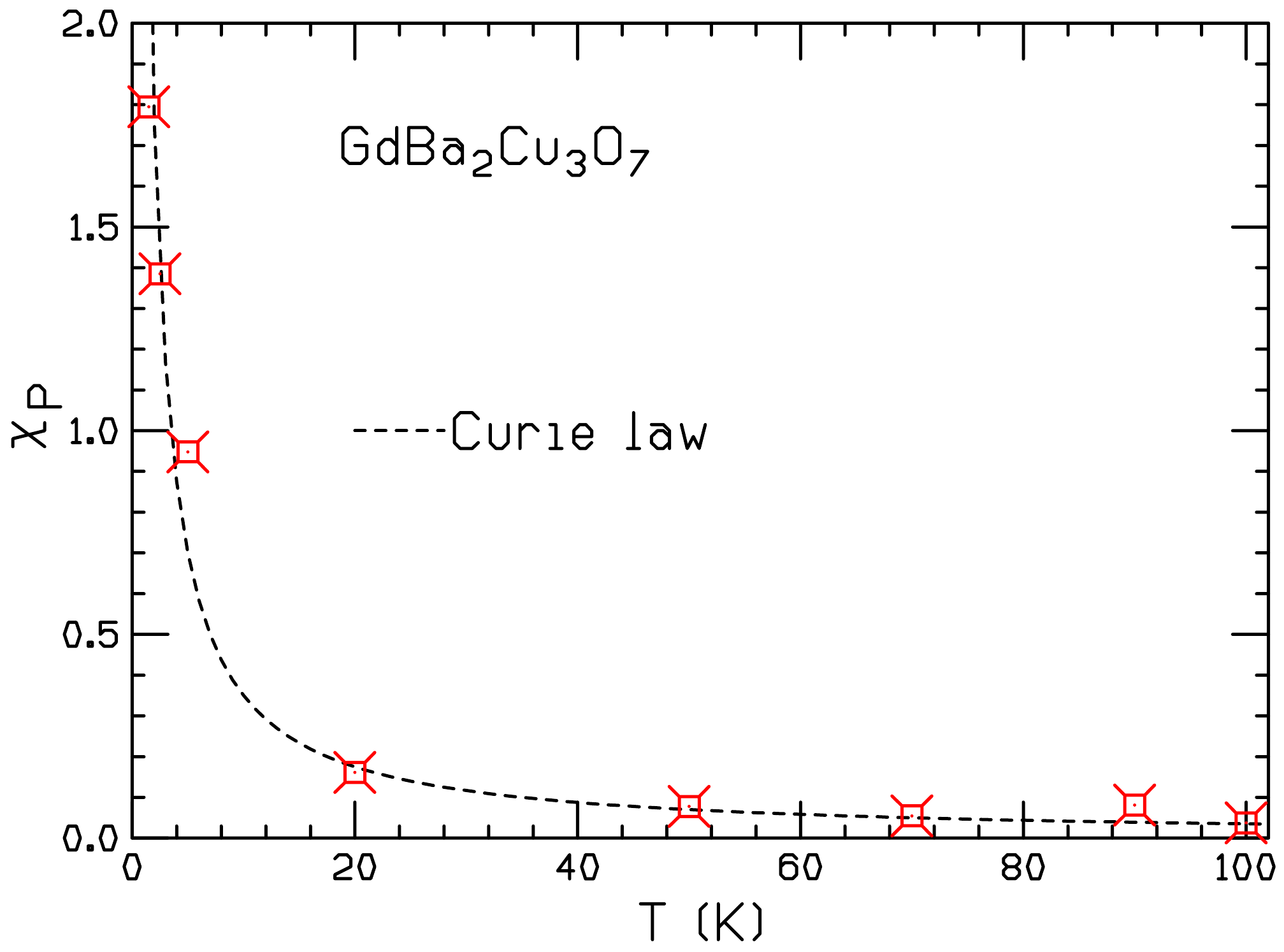}} 
\caption{Paramagnetic contribution to the susceptibility $\chi_P$ inferred from Fig.~3, in units 
emu/g/kG. 
The experimental data are the points.
}
\label{fig3}
\end{figure}

It can be seen that a simple Curie law fits the paramagnetic susceptibility data inferred from Fig.~3 quite well, in contrast to the
situation in Fig.~2. Therefore, it is completely plausible that the diamagnetic part of the response in Fig.~3 originates in
superconductivity while the paramagnetic part originates in the large magnetic moment of $Gd$ atoms  
that behave essentially as isolated magnetic impurities.
A more detailed analysis performed in Ref.~\cite{ref31} yields a Curie-Weiss temperature $\theta \sim -2\pm 1$~K.

We argue that the explanation proposed by the authors of \cite{sh3} for the behavior of the magnetization displayed in Fig.~1
is completely implausible.
The key point is that there would be no reason for 
the paramagnetic contribution to the susceptibility originating in  a background signal   to go to zero at essentially the same temperature at which the diamagnetic contribution from the sample goes to zero, as Fig.~2 indicates.
Instead, the fact that it does, indicates that the paramagnetic contribution and the diamagnetic contribution originate from the same physical object, presumably the sample.

We then have to ask: is there any other standard superconductor where such a behavior of the magnetic susceptibility 
below $T_c$ is observed? We define `standard superconductor' theoretically by superconductors
described in Tinkham's book \cite{tinkham}, and experimentally by superconductors belonging to any of the 
32 families of conventional, unconventional and possibly unconventional  superconductors surveyed in Ref.~\cite{specialissue}, that
cover all superconducting materials known in 2015.
In Ref.~\cite{specialissue}, materials were classified as `conventional' (classes C1 to C12), `unconventional' (classes U1 to U11) and
`possibly unconventional' (classes P1 to P9).

Among these superconducting material classes  there are the conventional magnetic superconductors (C10) and the
ferromagnetic superconductors (U7), that one could consider as prime candidates for exhibiting large
paramagnetic susceptibility below $T_c$ superposed to a diamagnetic response. 
One may also consider numerous other classes that exhibit magnetic phases nearby in the phase boundary 
such as Chevel phases (C9), borocarbides (P3), plutonium compounds (P4), heavy-fermion 
superconductors (U1), organic compounds (U2), cuprate superconductors (U3, U4), 
strontium ruthenate (U5),  cobalt oxyde hydrate (U8), iron-based superconductors (U10, U11). However, none of these materials exhibit the anomalous behavior of
magnetic susceptibility below $T_c$ exhibited by sulfur hydride shown in Fig.~1.

Regarding theory, Tinkham's book describes the phenomenology of the magnetic behavior of  standard type I and type II superconductors
expected from London theory and Ginzburg-Landau theory, irrespective of the particular microscopic mechanism giving rise
to superconductivity \cite{tinkham}. That physics does not  give rise to  the behavior of magnetization versus field shown in Fig.~1.
 
 \section{new physics or experimental artifact?}
 Forgetting for the moment about superconductivity, we know of no magnetic material  that 
 would show a behavior of its magnetic susceptibility like the one shown in Fig.~2.
 
 The magnetic response of materials originates in orbital magnetism or spin magnetism or a combination of both. Let us recall 
 the known behaviours:
 
 1) Temperature independent paramagnetic susceptibility is seen in metals due to the electron magnetic moment and Fermi statistics.
 
 2) Small temperature independent diamagnetic susceptibility is seen in insulators with atoms with closed shells.
 
 3) Temperature dependent paramagnetic susceptibility originates in localized magnetic moments or in the magnetic moment of
 itinerant electrons in magnetic metals,
 and results in Curie  or Curie-Weiss behavior with a Curie-Weiss temperature $\theta$. If $\theta$ is positive
 the material undergoes a transition to a ferromagnetic state at $T=\theta$. If $\theta$ is negative the material may undergo
 a transition to an antiferromagnetic state at a finite temperature.
 
 The magnetic susceptibility shown in Fig.~2 increases by a factor of over 60 from $200$~K to $50$~K, i.e. in a temperature range where
 the absolute temperature decreases by a factor of 4. We know of no other way that this could happen 
 other than the system undergoing a transition to a ferromagnetic state at temperature slightly below $50$~K
 (the fit shown in Fig.~2 assumed a Curie-Weiss temperature $\theta=47.5$~K). However 
 it is impossible to fit the susceptibility data at temperatures in-between $50$~K and $200$~K with such behavior, as seen in Fig.~2.
 
 Within mean field theory, the critical temperature for onset of ferromagnetic order in a system of magnetic moments
 interacting through a ferromagnetic coupling $J$ can be written in the form
 \beq
 T_c=\frac{zJ}{k_B}
 \eeq
 where z is the number of nearest neighbors to a moment and $k_B$ is Boltzmann constant. 
 The magnetic susceptibility is given by (Curie-Weiss law)
  \beq
 \chi_P(T)=\frac{\chi_0}{T/T_c(T)-1}
 \eeq
 with $\chi_0$ a constant.  To describe the behavior shown in Fig.~2 we would have to assume that $T_c$ in Eq.~(2) is itself
 a function of temperature, as indicated in Eq.~(2), smaller than $T$ and decreasing as $T$ decreases with increasingly smaller
 $(T-T_c(T))$, so as to give the approximately linear behavior shown in Fig.~2. 
 This could result from a temperature dependent
 $z(T)$ or $J(T)$ in Eq.~(1) decreasing as $T$ decreases. However  we know of no physical mechanism that would give rise to such behavior.  
 
 There have been suggestions in the literature that hydrogen or hydrogen-rich compounds under high pressure
 would undergo a transition to a magnetic phase at low temperatures rather than a superconducting
 phase \cite{hirsch, mazov}. Such models would possibly explain the large paramagnetic response seen at low temperatures,
 but are unlikely to yield the peculiar linear temperature dependence of the magnetic susceptibility shown in Fig.~2.

 Considering now the superconductivity, note that the magnetic field range in Fig.~1 is a very small fraction of the
 assumed upper critical field in this material, approximately $70$~T \cite{sh3}. Nonetheless, since the field values are likely higher 
 than $H_{c1}$, we assume  the system is in the mixed state,
 with normal vortex cores of radius $\xi$, the coherence length. and the magnetic field screened over a radius
 $\lambda_L$, the London penetration depth, around each vortex. 
The total magnetic field is given by 
 \beq
 B=H+4\pi M
 \eeq
 where $H$ is the applied field and $M$ the magnetization. 
 Since each vortex can hold exactly one quantum of flux $\phi_0$,
 as the temperature is lowered the number of vortices must increase to accommodate the extra
 magnetic flux resulting from the induced magnetization (in the range where the sample is in the mixed state).  
 This should cause the diamagnetic response to decrease
 as the temperature is lowered rather than increase rapidly as indicated by the data in Fig.~1. 
 
 In summary, we cannot imagine a plausible  explanation for the reported linear increase in paramagnetic response   as the temperature is lowered
 within the conventional understanding of magnetism in solids, and the joint increase in paramagnetic and diamagnetic
 response as the temperature is lowered shown in Fig.~1 appears to contradict everything we know about superconductivity in 
 standard superconductors. Furthermore  we have argued that it is not plausible that the paramagnetic response
 originates in a different physical region of the apparatus used in the experiment, given the fact that it goes
 essentially to zero at the same temperature where the diamagnetism disappears.
 
 We conclude that it is highly likely that the experimental data shown in Fig.~1 that were reported in  Ref.~\cite{sh3} are an experimental artifact
 and do not reflect real physics.
  
\section{other magnetic evidence for superconductivity in   hydrides under high pressure}

Since 2015, superconductivity in high pressure hydrides evidenced  by resistance measurements in diamond anvil cells has been reported in the following compounds: 
$SH_x$ up to $203K$ \cite{sh3,sh32}, $PH_x$ above 100K \cite{eremetsp}, 
$LaH_x$ at 250K \cite{eremetslah,zhaolah}, above 260K \cite{hemleylah} and above
550K \cite{hemleylah2}, $YH_x$   at 243K \cite{yttrium2,yttrium,yttriumdias}, $ThH_x$ at 161K \cite{thorium}, $PrH_x$ at 9K \cite{pr}, 
$LaYH_x$ at 253K \cite{layh10}, $CSH$ at room 
 temperature \cite{roomt}, $CeH_x$ above 120K \cite{ceh}, $SnH_x$ at 70K \cite{snh}, $BaH_x$ around 20K \cite{bah}, and $CaH_x$ around 215K \cite{cah,cah2}.
 That is 19 different reports on 12 different materials. These papers have been cited 2,450 times according to Google Scholar (2/24/21), with essentially
 all citing papers assuming that superconductivity in these materials is an established fact. Numerous news media have also informed the general public that
 superconductivity in these materials is an established fact.
 
 However, to establish that a material is a superconductor it is essential to show that it exhibits a response to magnetic fields expected for superconductors. Many of the papers cited above limit themselves to consider the effect of magnetic field on resistance. We will
 discuss that effect at the end of this section. However, concerning the $magnetic$ response to magnetic fields, 
we have reviewed the existing literature on high pressure superconductivity in hydrides looking for such information and  have found only 5 papers, on three different materials, namely:

\begin{itemize}

\item  The magnetization measurements on sulfur hydride  \cite{sh3} discussed  in the previous sections.

\item  Ac magnetic susceptibility measurements for $CSH$  \cite{roomt}.

\item  Ac magnetic susceptibility measurements for sulfur hydride \cite{magsh3}.

\item  Ac magnetic susceptibility measurements for $LaH_x$  \cite{maglah}.

\item  Nuclear resonant scattering (NRS) measurements for sulfur hydride \cite{nrs}.

    \end{itemize}
Let us consider each of these.

\subsection{Magnetization of  sulfur hydride}
As discussed in the previous sections, the magnetization versus magnetic field data given in \cite{sh3}  shown in Fig.~1 of this paper are not consistent with the behavior of standard superconductors and magnetic materials. The paper \cite{sh3} also plots magnetization versus temperature in its Fig.~4a, which we reproduce in Fig.~5, showing that it
increases sharply from approximately $ -4\times 10^{-7}$ emu to $ +7\times 10^{-7}$ emu as the temperature is raised across $200$~K.
 
%figure 5
\begin{figure} [h]
 \resizebox{5.5cm}{!}{\includegraphics[width=6cm]{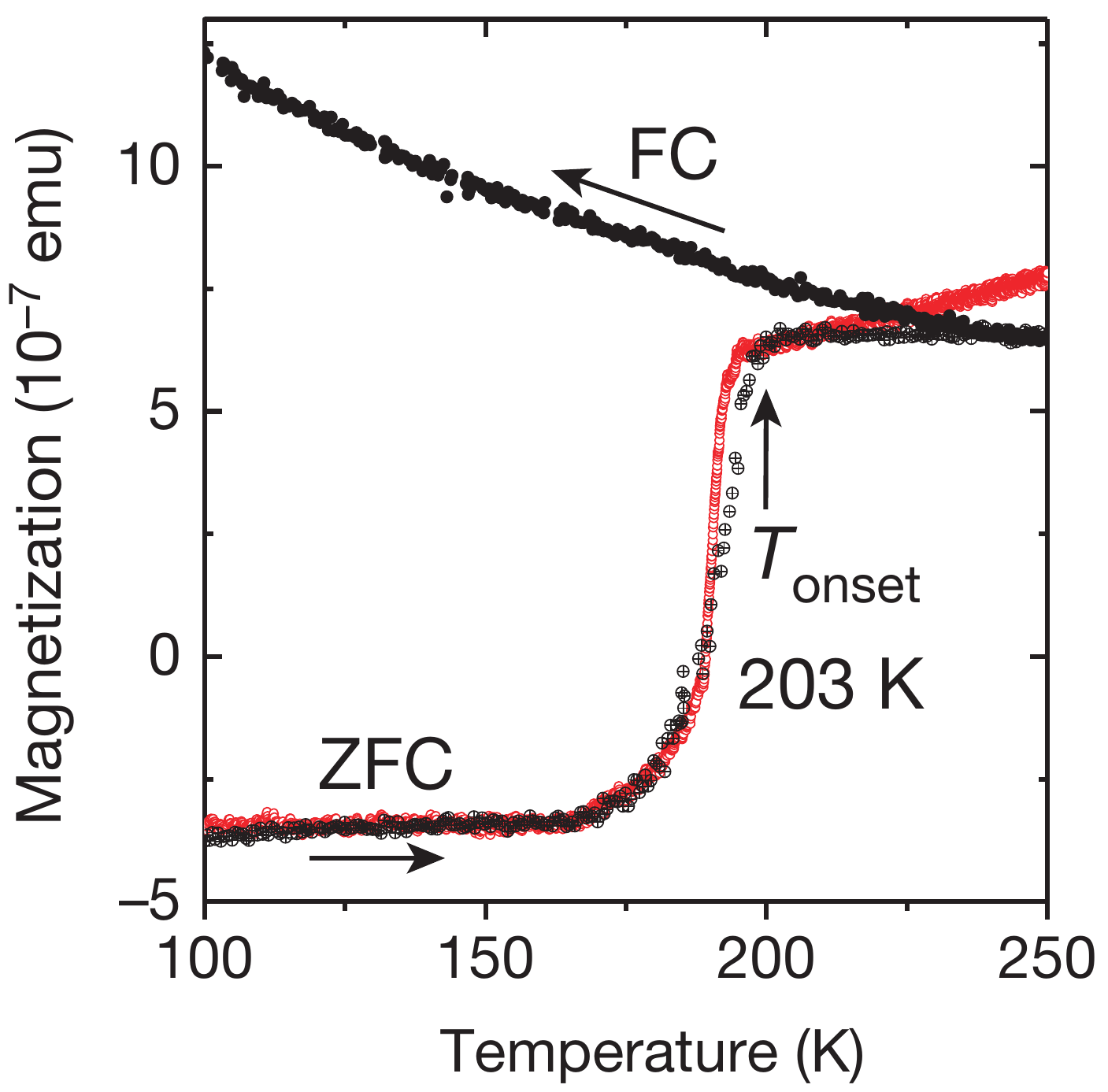}} 
\caption{Magnetization versus temperature reported in Ref.~\cite{sh3} (Fig.~4(a)), for field cooling (FC) and zero field cooling (ZFC),
or more accurately, field heating. The red dots are resistance data ``scaled and moved vertically to compare with the
magnetization data'' \cite{sh3}.
}
\label{fig3}
\end{figure}

\begin{figure*} []
 \resizebox{16.5cm}{!}{\includegraphics[width=6cm]{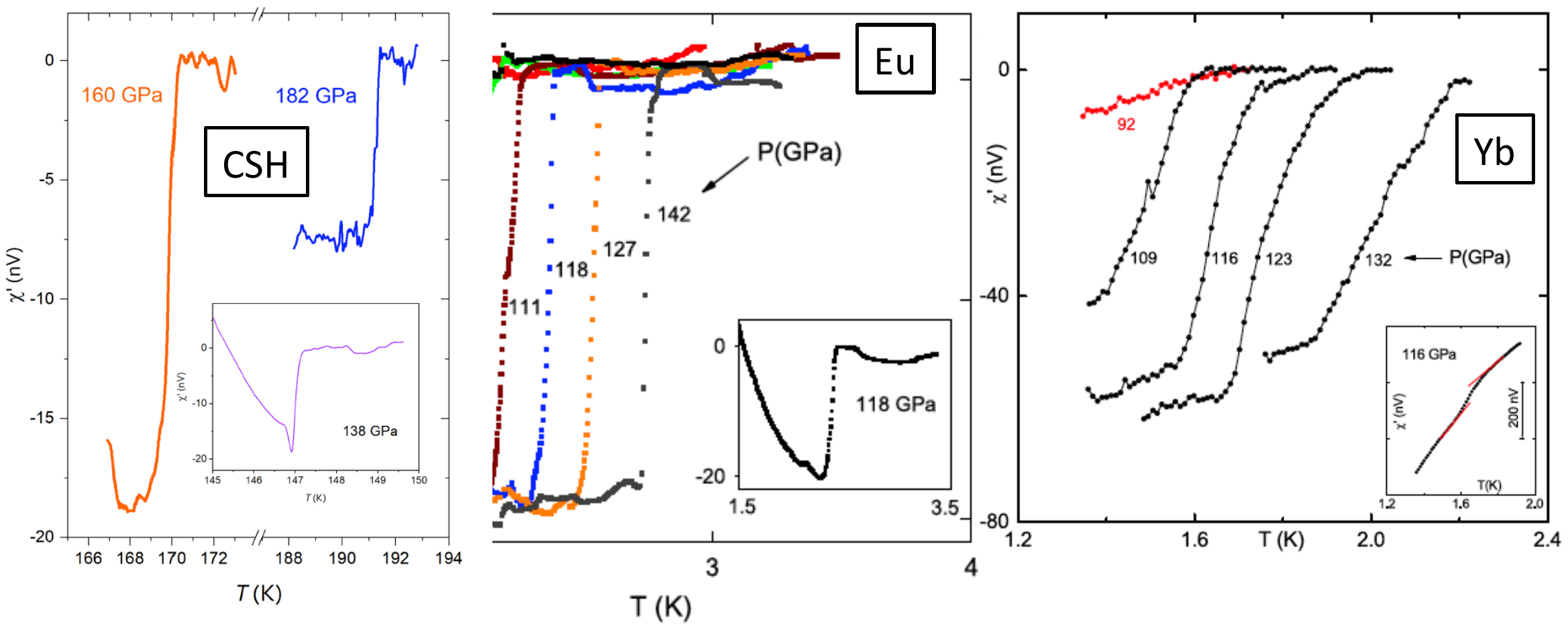}} 
\caption{Ac magnetic susceptibility for three materials under pressure in DAC's: 
CSH (\cite{roomt}), and the elements Eu (\cite{eu}) and Yb (\cite{yb}). The insets show the raw data, from which
the signal shown is extracted by substraction of a background signal.
}
\label{fig3}
\end{figure*}

However, in their extended data Figure~6 the authors show that there is a `background' magnetization signal that is two orders of magnitude larger than the signal shown in Fig.~5. As explained in the figure caption of extended data Figure~6, a background signal was subtracted in order to plot the data in Fig.~5.
However because there
is no precise way to tell what part of the raw signal is `background', it means there is no way to know whether the sample signal 
actually does change sign from negative to positive as the temperature is raised, as Fig.~5 purports to show. 
In fact, the measured data did not provide information that the signal below $200$~K is diamagnetic and above $200$~K paramagnetic, and therefore the authors' statement in
reference to their fig.~4(a)  ``a sharp transition from the diamagnetic to the paramagnetic
state (Fig.~4a) was observed for zero-field-cooled (ZFC) material'' is profoundly misleading.  The reality is that there was no evidence
from the measured data that the low temperature state of the sample was diamagnetic. 

The myth has persisted for the ensuing 5 years until the present. In a recent review paper \cite{review2} the authors (one of whom also coauthored \cite{sh3})
state in reference to this figure ``The ZFC curve demonstrates a rather sharp transition from the diamagnetic to
the paramagnetic state, which classified as a superconducting one. The onset temperature is $T_c = 203$~K.''

It should also be noted that the field cooled (FC) data shown in Fig.~5  show absolutely no signature of anything happening around $200$~K. This is also in contradiction with the behavior of
standard superconductors, where typically a smaller but non-zero signal of the transition is seen also in FC
measurements.

\subsection{Ac magnetic susceptibility for CSH}

Reference \cite{roomt} reports ac magnetic susceptibility measurements on a carbonaceous sulfur hydride (CSH)
that  show sharp drops at  temperatures close  to where the resistance versus temperature
curves show drops. An example from Ref.~\cite{roomt} is shown on the left panel of Fig.~6.
We will compare this behavior with other such data for materials in diamond anvil cells under high pressure, namely
the elements Eu \cite{eu} (middle panel) and Yb \cite{yb} (right panel).

As pointed out in Ref. \cite{eumine}   there is a problem with the data for both CSH and Eu. It is well known  \cite{hamlinthesis} that 
 in these experiments
there is a large background signal that needs to be substracted from the raw data to extract the susceptibility of the sample.
In the right panel of Fig.~6, the inset shows raw data obtained for Yb under pressure by Song {\it et al.} \cite{yb}, and the right panel shows
the extracted signal. The raw data reflect a smooth linear background, and superposed to it a small drop that is
plotted in the main right panel. These measurements together with the resistance measurements reported in  \cite{yb}
give clear and unambiguous evidence that Yb becomes superconducting in these experiments.

Instead, the raw data shown in the insets of the left and middle panels of Fig.~6, for both CSH and the element Eu show features that
are inconsistent with superconductivity, namely: (i) the fact that the raw signal is flat for temperatures above the drop, and (ii) the fact 
that the raw signal rises steeply for temperatures below the drop. These data are qualitatively different from the featureless linear background
seen in the right panel of Fig.~6 which is typical for these measurements. As discussed in Ref.~\cite{eumine},
this  indicates that both the data for CSH and Eu are either (a)  an experimental artifact and do not reflect real physics, or (b) 
an indication that a magnetic transition instead of a superconducting transition is taking place,  which could account for the
steep rise in the raw susceptibility data at lower temperature. The initial drop could be attributed to the sample
becoming more conducting as it develops magnetic order.

As noted  in Ref.~\cite{eumine}, the measurements of the left and middle panels of Fig.~6 were performed with the
same equipment and by  the same researcher, which supports  the possibility that they could be due to an experimental artifact. 

\subsection{Ac magnetic susceptibility for sulfur hydride}

Reference \cite{magsh3}  (see also \cite{semenok}) reports results of ac magnetic susceptibility measurements for sulfur hydride, that look qualitatively similar
to the data for CSH reported in \cite{roomt}, left panel of Fig.~6. However, Ref.~\cite{magsh3} does not give essential  information to allow the reader
to evaluate the significance of the data presented. It states 
``The protocol of the magnetic-susceptibility technique for measuring the superconductivity is described in previous works''
and cites three references, \cite{ref1,ref2,ref3}. However, it does not tell the reader 
what frequency was used in the measurements,
which is essential in order to understand the significance of the amplitude of the signal reported
(the frequencies used in Refs.~\cite{ref1,ref2,ref3} are all different, as are the detailed procedures).
Nor does Ref.~\cite{magsh3} give any information on the background signal, nor does it inform the reader that a subtraction of background signal
was performed, nor does it provide any raw data. 

We have attempted to obtain additional information to evaluate the significance of these data by contacting the authors 
of Ref.~\cite{magsh3} but did not receive any response. Given the limited information in the published paper, we argue that it cannot be
taken as clear evidence for a superconducting transition in sulfur hydride.

%figure 7
\begin{figure} [t]
 \resizebox{8.5cm}{!}{\includegraphics[width=6cm]{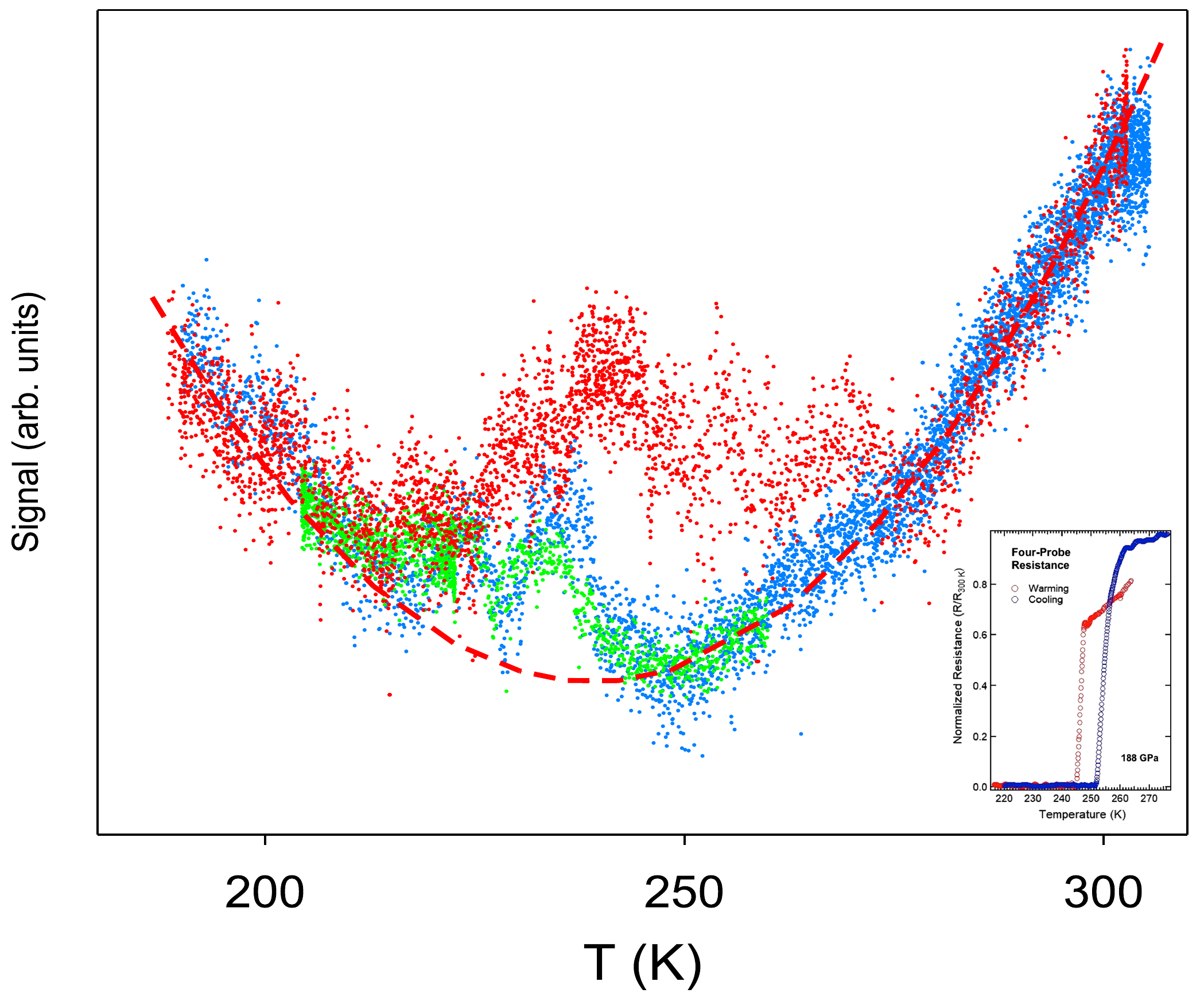}} 
\caption{ Magnetic response signals for $LaH_x$ from Ref.~\cite{maglah}. The blue, green and red points correspond to
pressures 164GPa, 169GPa and 180GPa. The assumed background is shown by a red dashed line \cite{maglah}.
The inset shows resistance versus temperature results for this material, from Ref.~\cite{hemleylah}.
}
\label{fig3}
\end{figure}

\subsection{Ac magnetic susceptibility for lanthanum hydride}

Reference \cite{maglah} reports ac magnetic susceptibility measurements on $LaH_x$, using 
a  low-frequency modulation technique discussed in Refs.~\cite{ref1,ref2}. The data are shown in Fig.~7.
This technique was $not$ used in Ref.~\cite{magsh3}, despite the ambiguous statement in Ref.~\cite{magsh3}  quoted above
that suggests otherwise. With this technique the evidence for superconductivity is detected as a peak in the signal rather than a drop.
Fig.~7 shows  very broad features, much broader than the width of
the resistive transitions reported in \cite{eremetslah,hemleylah, zhaolah} for this compound (see inset in Fig.~7). Note that their amplitude
relative to the background cannot be ascertained since no vertical scale is given. 
Ref.~\cite{maglah} states
``We were able to detect a measurable signal, however, the estimated
sample size proved to be very small'',  and suggests the possible existence of multiple phases 
and pressure gradients to account for the broad signals. 
Note in contrast that in sulfur hydride the width of the transition was reported to be the same in magnetic and resistive
measurements (Fig.~5, black points for ZFC and red points for resistance).

\subsection{Nuclear resonant scattering for sulfur hydride}

Reference \cite{nrs} reports results of a nuclear resonance scattering (NRS) experiment as evidence that 
that sulfur hydride expels magnetic fields.
In Ref.~\cite{hm3} we have carefully analyzed the data presented in Ref.~\cite{nrs}.
Fig.~8 shows schematically the magnetic field configuration that Ref.~\cite{nrs} claims is detected in their  NRS experiment, proving that the
sample excludes magnetic fields. However, if this was true it would imply, given that the applied magnetic field was $0.68T$,  that the magnetic field at the edge of the sample was $2.5$~T \cite{hm3}, hence that $H_{c1}$, the lower critical
field of this material, is larger than $2.5$~T at temperatures  below $50$~K, where it is claimed that the field is totally excluded \cite{nrs}. 
This value of $H_{c1}$ differs from the lower critical field for this material estimated in Ref.~\cite{sh3} from the magnetization data,
$H_{c1}\sim 0.003T$, 
{\it by three orders of magnitude}. It is also more than two  orders of magnitude larger than any lower critical field known for any other
superconductor, and incompatible with other properties of the material as discussed in \cite{hm3}.

\begin{figure} [h]
 \resizebox{7.5cm}{!}{\includegraphics[width=6cm]{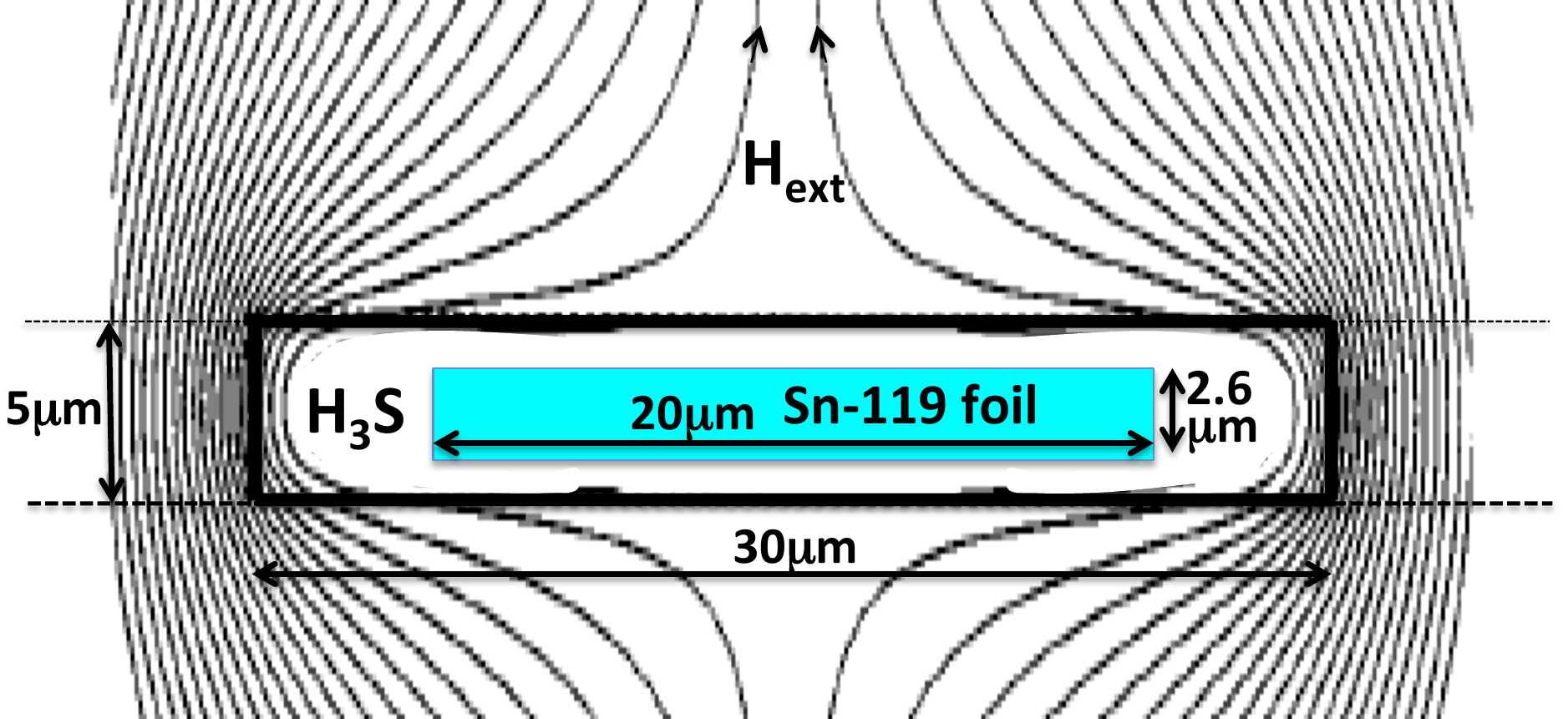}} 
\caption{Geometry of the NRS experiment with magnetic  field perpendicular
to the sensor film \cite{nrs,hm3}. The applied magnetic field was $0.68T$.
}
\label{fig3}
\end{figure}

It should also be noted that such an experiment to detect the Meissner effect using nuclear resonance scattering
has never been used, either before or after the experiment reported in \cite{nrs}, to detect the Meissner effect for this
or any other superconducting material according to the published literature.
\newline
\newline

The data surveyed above comprise the totality of measurements of the magnetic response of hydrides under high pressure to
applied magnetic fields that exist in the scientific literature to date. We argue that they do not provide strong evidence for the existence of superconductivity in these materials, for the variety of reasons discussed above.

\subsection{Effect of magnetic field on resistance}
Most of the 19 papers cited above that report the finding of high temperature superconductivity in hydrides measured the effect of magnetic field
on resistance, and found that a large magnetic field lowers the apparent transition temperature. From these measurements
they extract values of the upper critical magnetic field $H_{c2}$,   in the range approximately  between
$50T$ and $150T$.

However, we have pointed out in Refs.~\cite{hm,hm2} that the resistance in the presence of magnetic field that was found often does not
follow the behavior expected for normal superconductors, which is that the resistive transition broadens under application of
a magnetic field. This led us to conclude \cite{hm2}   that either these materials are ``nonstandard superconductors'',
with properties markedly different from those of standard conventional and unconventional superconductors, or more likely that they
are not superconductors. If the latter, the reason that the apparent $T_c$ shifts downward when a magnetic field is applied would
not be related to superconductivity.

To test this hypothesis, we propose  that the effect of magnetic field on resistance {\it as a function of the orientation of the
magnetic field with respect to the sample} should be explored. To our knowledge, all  the experiments performed to date 
that measured resistance applied the
magnetic field in  the direction perpendicular to the plane of the sample; this plane is defined as perpendicular to the short dimension
of the sample, which is parallel to the applied pressure. As shown in Fig.~8, the plane is horizontal. So in particular we suggest that the experiments should be
repeated with the magnetic field applied in the plane of the sample. 

We note that the lattice structures that have been hypothesized for these compounds are isotropic \cite{review1,review2,maglah}
and the pressure is assumed to be hydrostatic \cite{eremetslah,eremets2016}, so that
the upper critical field $H_{c2}$ should be the same in directions perpendicular and parallel to the plane of the sample.
Therefore, if the effect of the magnetic field on resistance is due to the onset of superconductivity, 
the temperature at which the resistance begins to drop should be $independent$ of the orientation of applied magnetic field.
The resistance below that temperature on the other hand may have some
variation with orientation of magnetic field perpendicular and parallel to the plane as well as for the   two orthogonal orientations possible within the plane.

Instead, if the drop of resistance is not due to superconductivity but for example to the action of the magnetic Lorentz force on normal charge
carriers giving rise to `geometrical magnetoresistance' \cite{mr1,mr2} we would expect a much smaller effect of magnetic field parallel 
versus perpendicular to the plane. Or, if the drop in resistance is related to the system developing magnetic order \cite{hirsch,mazov}
in some  particular direction, the effect of an applied magnetic field would be expected to depend on its orientation.

\section{discussion}
Electrical resistivity  in materials that are not superconducting 
covers a range from about  $10^{-8}\Omega m$ to $10^{20} \Omega m$, that is,  28 orders of magnitude.
For that reason a drop of two or three orders of magnitude in the resistance of a material as a function of temperature or pressure,
as typically reported for the hydrides,  
could  be due to many reasons {\it other than} superconductivity. It should also be noted  that papers in this field almost always show $resistance$
rather than $resistivity$ data due to the difficulty of obtaining reliable information on samples' dimensions. This  precludes
the possibility of evaluating the significance of the magnitude of the resistivity in these materials.
It should also be recalled that throughout the history of superconductivity there have been many reports of
`USO's, ``unidentified superconducting objects,'' based on observations of resistance drops, that were
never confirmed \cite{uso}.
Unlike what has been the tradition in the field of   superconducting materials, for the particular
class of hydrides under high pressure even  drops of resistance versus temperature by less than an order of magnitude
are interpreted as unmistakable signs of a superconducting transition \cite{sh32,snh,zhaolah}.

To establish that a material is a superconductor it is essential
to show that it exhibits the characteristic behavior in external magnetic fields that only superconducting materials do.
As we have shown in this paper, such evidence for the hydrides is scant and ambiguous at best.
Furthermore it is incomprehensible that such measurements once performed are not repeated in the same or in different labs with 
the same or higher
accuracy to either confirm or rule out the initial conclusions. 
The 2015 measurements of magnetization in sulfur hydride
discussed in this paper required the development of a new high pressure technique,
a specially designed miniature nonmagnetic DAC cell made of Cu-Ti alloy that could accomodate a SQUID magnetometer \cite{drozdov,review1}. 
Yet no results using this sophisticated apparatus other than the ones shown in Fig.~1, neither for sulfur hydride nor for any
other hydride, have been reported in the ensuing 
6 years to the present, during which the number of supposedly superconducting hydrides  
mushroomed from 1 to 11. The  nuclear  resonant study that supposedly proved 
the Meissner effect for sulfur hydride in 2016 \cite{nrs}, discussed in Sect. IV and in Ref.~\cite{hm3}, has never been repeated for any other superconductor,
hydride or otherwise.

Based on the analysis of the magnetic evidence discussed in this paper, together with the anomalous 
resistance versus temperature behavior in several hydrides discussed in our recent work \cite{hm,hm2}  and
that of others \cite{dc}, we conclude that the existence of high temperature superconductivity in hydrides
under high pressure is $not$ an established fact. This conclusion is in contradiction with the general consensus.
So we may ask: why have most scientists, both those that work in the field and those that watch it from nearby, 
uncritically accepted the existence of
high temperature superconductivity in pressurized hydrides as a   fact, on the basis of flimsy experimental evidence?

Examination of the literature in this field suggests  the reason for this anomalous state of affairs: the field is entirely  driven by theory.
Essentially  every experimental paper reporting measurements interpreted as superconductivity in a hydride starts with
a lengthy exposition of why the conventional BCS-Eliashberg-electron-phonon theory of superconductivity makes the existence
of high temperature superconductivity in the material inevitable and unavoidable.
Sometimes experimental evidence for superconductivity is only presented towards the end of the paper, almost as an afterthought \cite{pr}.
In cases where the experiment doesn't find what   the theory predicted, the experiment is not interpreted as casting doubt into the
applicability of the theoretical framework used, instead a new twist is added to the theory to agree with the experimental observation,
or  it is concluded that   `the sample  made a mistake' \cite{euh0,euh}. 
Purely theoretical papers are written and titled in such a way that they appear to report the finding of a  
new  superconducting material in real life   when in actual fact it was only found in a computer   \cite{tih,hsz,labh8}.

{\it If}  BCS-Eliashberg-electron-phonon theory is the correct theory to describe the superconductivity of a large set of materials in nature,
the so-called conventional superconductors, 
such an approach may be justified. However, it should be recalled that 
historically, conventional BCS-Eliashberg theory coupled with density-functional band structure calculations
has $not$ been an effective tool to predict and  find new superconductors \cite{matthias}. Are we to believe that among the 32 classes of superconducting
materials surveyed in \cite{specialissue}, the theory is singularly effective in  predicting new superconducting materials \cite{review1,review2,theory1,theory2} in class 
C7, compressed hydrogen-rich materials \cite{c7}, and not  in the other 31 classes? Why would that be the case?

For the moment, unless or until the high temperature superconductivity of hydrides under pressure
is convincingly established through experiments that reproducibly measure the magnetic response expected for superconductors,
we suggest that an alternative possibility should be considered:   that the high pressure hydrides may in fact {\it not be} high temperature superconductors; hence that   BCS-Eliashberg  theory may not be an effective tool to predict superconducting materials in
{\it any} of the 32 classes of materials surveyed in \cite{specialissue};  hence that specifically  the very definite prediction of the theory that high  temperature superconductivity exists in pressurized hydrides  may   not be correct.
If that is proven to be the case, if such a compelling prediction of the theory is proven wrong, the possibility   that
BCS-Eliashberg-electron-phonon theory may  not correctly describe the physics of superconductivity of $any$ superconducting material, as was proposed
by one of us in Refs.~\cite{validity,book}, should become compelling.

\begin{acknowledgments}
 JEH is grateful to V. Struzhkin for a  stimulating discussion.
FM  was supported in part by the Natural Sciences and Engineering
Research Council of Canada (NSERC).  

\end{acknowledgments}

\end{document}